\documentclass[10pt,twoside]{hsqcd}
\usepackage{epsf,amsmath}
\usepackage{graphicx}

\begin{document}

\title{QCD Sum Rules and 1/$N_c$ expansion:
	On the low energy dominance and
	separation of scattering backgrounds
}

\author{\underline{Toru Kojo}$^1$ and Daisuke Jido$^2$\\ \\
$^1$RBRC, Brookhaven National Laboratory,
Upton, New York 11973, USA \\
$^2$YITP, Kyoto University,
          Kyoto 606-8502, Japan \\
E-mail: torujj@quark.phy.bnl.gov}

\maketitle

\begin{abstract}
\noindent
In multiquark correlator analyses
with $1/N_c$ classifications,
it is possible to 
separate the scattering background and
to justify the factorization of condensates,
which allows us to achieve an {\it isolated peak saturation} in the
QCD sum rules for multiquark currents.
Then we can extract leading $N_c$ properties of the ground state.
An application to the $\sigma$ meson is demonstrated.
\end{abstract}



\markboth{\large \sl \underline{T. Kojo} \& D. Jido
\hspace*{2cm} HSQCD 2008} {\large \sl \hspace*{1cm} TEMPLATE FOR THE
HSQCD 2008 PROCEEDINGS}
\section{QCD sum rules for multiquark correlators: Difficulties}
The multiquark states have 
been intriguing subjects of researches
to study the infrared aspects of QCD
beyond those manifested in the usual hadron spectra.
One of the conventional approaches from QCD
is to analyze correlators of multiquark currents,
which is related to the hadronic spectral functions through
the dispersion relation.
Compared to analyses for usual meson and baryon correlators,
the multiquark cases are expected to have the following difficulties:
(i) The complications associated with the many-body system,
which make the calculation of correlators much involved.
(ii) The decays into usual hadrons, whose backgrounds contaminate
the signal from multiquark states.
We will show that the use of $1/N_c$ expansion \cite{tHooft, Witten}
greatly reduces both difficulties, and, more importantly,
enables to make logically clear statements \cite{KJ2}.

We perform the correlator analyses using the Borel transformed
QCD sum rules (QSR) for the hadronic correlators 
$\int dx e^{iqx} \langle T \hat{J}(x)\hat{J}^{\dag}(0) \rangle$ 
\cite{Shifman}:
\begin{eqnarray}
\sum_n C_n(M^2;\mu^2) \langle \hat{O}_n(\mu^2) \rangle
= \int_0^{\infty} \!\! ds\ e^{-s/M^2} \frac{1}{\pi} {\rm Im}\Pi^h(s),
\label{QSR}
\end{eqnarray}
where LHS is calculated by the operator product expansion (OPE)
and RHS is the integral of the hadronic spectrum.
Here $M$ characterizes
the scale of momentum flow between interpolating fields, and
$\mu$ is factorization scale separating high energy calculable part
and low energy condenstate part.

Usually we are interested in the ground state properties of
RHS in eq.(\ref{QSR}), and they can be investigated 
by taking small $M^2$ due to the exponential factor $e^{-s/M^2}$.
But we can not take arbitrarily small $M^2$ value
since the OPE is basically expansion of
$\langle\hat{O_d}\rangle/M^{d} \sim (\Lambda_{{\rm QCD}}/M)^d$ 
and is violated in small $M^2$ region.
Thus we must take intermediate region
where the integral is well-saturated by ground state contribution
and the OPE is well-converged.
These are quantified as:
pole contribution $\ge$ 50\% of the total 
\cite{KHJ,KJ} (for upperbound $M^2_{max}$)
and highest dimension terms $\le$ 10\% of whole OPE
(for lowerbound $M^2_{min}$).
Practically, without these $M^2$ constraints,
we will be stuck with the {\it pseudo-peak} artifacts
leading to wrong conclusions, as discussed in \cite{KJ}.

Keeping these constraints in mind,
let us consider how the first difficutly (i) is manifested
in the QSR analyses.
The multiquark correlators involve many quark lines,
sharing only small fraction of the momentum with a scale 
relevant to the multiquark states. 
As a result, we have to incorpolate more non-perturbative diagrams 
in the OPE calculations than usual hadron analyses,
to include relevant low energy correlations. 
Otherwise, we can not find the $M^2$ window due to 
the lack of sufficient low energy contributions.

Here we shall add one remark about the low energy dominance.
In the case of multiquark correlator analyses,
it is frequently said that 
the pole-dominace is spoiled
by large continuum (or high energy part of spectrum) contributions
peculiar to multiquark correlators \cite{KJ}.
This is a somewhat misleading expression.
A real problem in the early studies is
the lack of low energy correlation.
In fact, in certain tetraquark correlators,
inclusion of higher dimension terms enables to take sufficiently
small $M^2$ value, achieving the low energy dominance not much worse than
usual meson sum rules, and even better than baryon sum rules \cite{KJ}.
Thus we conclude that a relevant problem
is numerical ambiguities in values of higher dimension condensates.

The second difficulty (ii) peculiar to multiquark
cases arises as follows:
even if we take into account the sufficient
low energy correlations, 
they might be contributions from scattering backgrounds
which are ground states in most of multiquark correlators.
This prevents us from investigating the multiquark states
of our interests.

Both difficulties seem profound.
However, as we will show,
leading $N_c$ properties of multiquark states
are tractable
and, in principle, the corrections
to them can be treated step by step
with identifying origins of contaminations.
\section{Factorization of condensates
	and separation of background based on $1/N_c$}
In this section, we discuss
how useful $1/N_c$ expansion is in the application
of the QSR.
The first merit is presented in the evaluation of 
the higher dimension condensates.
It is well-known that in large $N_c$
they can be factorized into
the products of known condensates,
$\langle \bar q q \rangle$, $\langle G^2 \rangle$, and
$\langle \bar q g_s \sigma G q \rangle$.
Thus we can extend the OPE calculations to 
sufficiently higher dimension terms needed to
include relevant low energy contributions.
Here we must make one assumption
on the $N_c$ scaling of these low dimension condensates.
Since $\langle O \rangle=$($\langle \bar{q}q\rangle$,
$\langle \alpha_s G^2 \rangle$, 
$\langle \bar q g_s \sigma G q \rangle$)
are known to be $O(N_c)$,
we simply assume
$\langle O \rangle |_{N_c} = \langle O \rangle N_c/3$.

Next, we discuss how to separate
the scattering background from the multiquark correlators.
The point is that
$N_c$ counting of the quark-gluon graphs
is directly related to
the qualitative picture of hadronic states.
We will show that with taking appropriate interpolating fields,
the multiquark state and background
can be assigned in the different order of $N_c$,
and their mixing occurs only in the higher order of $1/N_c$
than those studied in this work.
Then we can concentrate on the quark-gluon graphs
which are saturated by the {\it isolated} poles
which reflect leading $N_c$ properties of multiquark states.

To make discussions concrete,
we consider the $\sigma$ meson with
$I=J=0$ as a candidate of the tetraquark (4q) state.
The 4q operators with the $\sigma$ quantum number
are given (assuming the ideal mixing for the $\sigma$ meson)
by $J_{MM}(x) = \sum_{F=1}^3 J_{M}^{F}(x) J_{M}^{F}(x)$
as products of meson operators
$J_{M}^{F}=\bar{q} \tau_F \Gamma_M q$,
where Dirac matrix $\Gamma_M =(1, \gamma_{\mu})$
and $\tau_F\ (F=1,2,3)$ are the Pauli matrices 
acting on $q=(u,d)^T$.

\begin{figure}[b]
\vspace{-0.4cm}
\begin{center}
\hspace{0.7cm}
\includegraphics[width=13.0cm, height=2.4cm]{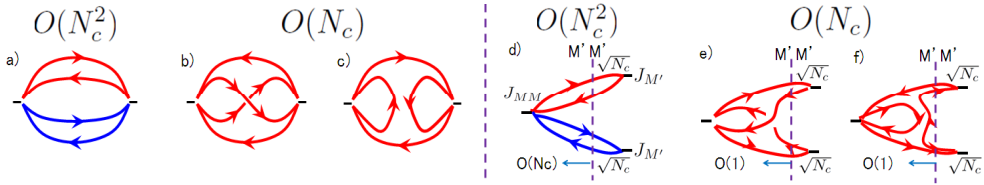} 
\vspace{-0.3cm}
\caption{$O(N_c^2)$ and $O(N_c)$ quark-gluon diagrams  
for 2 and 3 point correlators.}
\vspace{-0.0cm}
\label{fig:2pointgraph}
\end{center}
\end{figure}

Let us start the $1/N_c$ classifications of
quark-gluon graphs in the 2-point correlators, graphs (a)-(c) 
in Fig.\ref{fig:2pointgraph}, 
where planar-gluon lines are not explicitly drawn.
The leading $N_c$ diagrams start from
Fig.\ref{fig:2pointgraph} a), which
include only 2 planar loops and thus
reflect irrelevant free 2-meson scattering states.
This indicates that
the studies of 4q components
require systematic steps
beyond the leading $N_{c}$.
Thus we must proceed to the next leading order of $1/N_{c}$,
$O(N_c)$ diagrams
which could include the multiquark states.

For systematic classifications,
it is useful to first investigate the overlap strength
of the operator $J_{MM}$ with hadronic states by $1/N_c$,
then to use them in
the classification of subleading diagrams of 2-point correlators.
For this purpose, we consider 
the 3-point correlator among $J_{MM}$ 
and two separated $J_{M'}$
(Fig.\ref{fig:2pointgraph}, d-f).
Simple calculations
indicate that the overlap strength of the 4q field 
with 2 meson states is
$\langle 0|J_{MM}|M'M'\rangle = O(N_c)\delta_{MM'} + O(1)+....$
since the leading order diagrams are $O(N_c^2)$ ($O(N_c)$) for $M=M'$
($M \neq M'$) and
the overlap strength of $J_M'$ with 2q meson state $|M'\rangle$ 
is $O(N_{c}^{1/2}$) \cite{Witten}.

Now we can classify
the hadronic states in
2-point correlators $\langle J_{MM} J_{M'M'} \rangle$ 
based on $1/N_{c}$ (See, Fig.\ref{fig:2pointgraph}, a-c):
(a) If $M=M'$, $O(N_c^2)$ quark-gluon graphs include only the free
2M scattering states in the region $E \ge 2m_M$.
If $M \neq M'$, the contributions from these quark-gluon diagrams vanish,
indicating the absence of free 2 meson scattering states.
(b) $O(N_c)$ graphs include the 2M or 2M' scattering
and could include 4q poles.

Note that
if $M,M'$ are different from pseudoscalar or axial vector,
2$\pi$ scattering intermediate states 
are not included up to $O(N_c)$ diagrams.
Interactions are needed to transfer initial states
into the $2\pi$ intermediate states,
but such interactions are suppressed by $1/N_c$.
Then the resonance peaks (if exist) {\it below} $2m_{M}$
are {\it isolated and have no width}
since the decay channels are absent at this order.
Therefore,
now we can reduce the $\sigma$ spectrum in the 4q correlator
into peak(s) plus continuum  
{\it if we retain only diagrams up to $O(N_c)$}
for the appropriate currents.
We will study these cases.

This separate investigation of the $O(N_c^2)$
and $O(N_c)$ parts enables to perform
the step by step analyses for each hadronic state.
We relate the OPE, {\it term by term of $1/N_c$},
to the {\it integral} of the hadronic 
spectral function through the
dispersion relation:
\begin{eqnarray}
{\rm \Pi}^{ope}_{N_c^n}(-Q^2) 
= \int_0^{\infty}\hspace{-0.2cm}
ds\ \frac{1}{\pi}\frac{ {\rm Im \Pi}^h_{N_c^n}(s) }{s+Q^2}
\ \ (n=2,1).
\label{disp}
\end{eqnarray}
The ground state in the reduced
$O(N_c)$ spectra ${\rm Im \Pi}_{N_c}^{h}(s)$
can be described as the sharp peak 
due to the absence of the decay channels.
Applying the usual quark-hadron duality
to the higher excited states,
$\pi{\rm Im \Pi}^{h}_{N_c} (s) = \lambda^2 \delta(s-m_h^2)
+ \theta(s-s_{th}) \pi {\rm Im \Pi}_{N_c}^{ope} (s)$,
and taking the moment of Borel transform of Eq.(\ref{disp}),
we can express the effective mass as
\begin{eqnarray} 
m_h^2(M^2;s_{th}) \equiv \frac{ \int_0^{s_{th}} \!ds
\ e^{-s/M^2 }s\ {\rm Im \Pi}_{N_c}^{ope} (s) }
 { \int_0^{s_{th}} \!ds\ e^{-s/M^2 }{\rm Im \Pi}_{N_c}^{ope} (s) }.
 \label{eq:peakmass}
\end{eqnarray}
$s_{th}$ can be uniquely fixed to satisfy the
least sensitivity \cite{Shifman} of the expression 
(\ref{eq:peakmass})
against the variation of $M$,
since the physical peak should not depend on the
artificial expansion parameter $M$.
This criterion is justified only when the peak
is very narrow,
and our $1/N_c$ reduction of spectra 
is essential for its application
to allow the QSR framework to determine all physical parameters
($m_h, \lambda, s_{th}$)
in a self-contained way.

\section{Borel analyses for the reduced hadron spectra}
\begin{figure}[b]
\vspace{-0.5cm}
  \begin{center}
    \begin{tabular}{ccc}
      \resizebox{60mm}{!}{\includegraphics{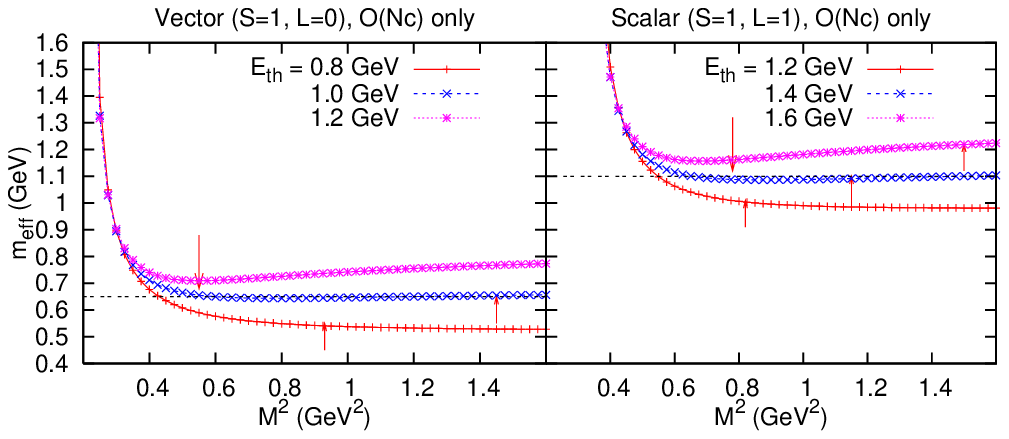}} &
      \resizebox{35mm}{!}{\includegraphics{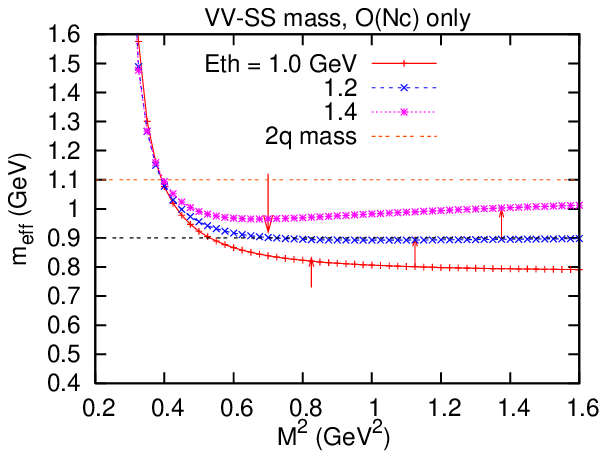}} &
	\hspace{-0.5cm}
      \resizebox{35mm}{!}{\includegraphics{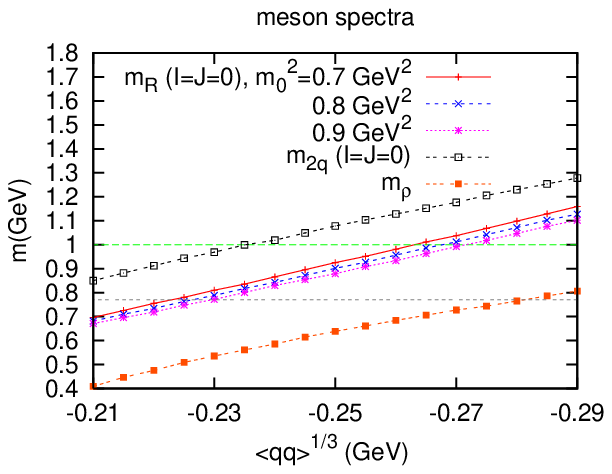}} \\
    \end{tabular}
    \caption[*]{The large $N_c$ 2q correlator results for
   the scalar and vector mesons (left),
	the $O(N_c)$ part of the 4q correlators (middle),
 and their mass relation for the various condensate values
(right).}
	\label{fig:borelfig}
  \end{center}
\end{figure}
In this section, we show the results of Borel analyses.
We concentrate on the $O(N_c)$ results of
the off diagonal correlator $\langle J_{VV}J^{\dag}_{SS} \rangle$,
whose leading order is $O(N_c)$
thus without the factorization violations
at the $O(N_c)$ OPE. 
The OPE is carried out up to dimension 12.
The numerical analyses are performed with
the values for $N_c=3$ case,
$\alpha_s({\rm 1GeV}) =0.4$,
 $\langle \alpha_s G^2/\pi \rangle = (0.33\ {\rm GeV})^4$,
$\langle \bar{q}q\rangle=-(0.25 \pm 0.03\ {\rm GeV})^3$,
and 
$m_0^2 = \langle \bar q g_s \sigma G q \rangle/ 
\langle \bar{q}q\rangle
= (0.8 \pm 0.1)\ {\rm GeV^2}$, 
respectively.
Most results shown below will use
the central value.

First we show in the left panel of Fig.\ref{fig:borelfig}
the results of
the large $N_c$ 2q correlators (expanded up to dimension 6)
for the vector meson as a reference and the scalar meson as the 2q state
in the $\sigma$ meson.
The downarrow (upperarrow) indicates the values of 
$M^2_{min}$ ($M^2_{max}(s_{th})$).
Following the $E_{th}$ ($\equiv \sqrt{ s_{th} }$) fixing criterion,
we fix $E_{th}$ to 1.0 (1.4) GeV for vector (scalar) mesons,
and determine the mass as 0.65 (1.10) GeV.

Now we turn to the $O(N_c)$ part of
the 4q correlator,
$\langle J_{VV}J^{\dag}_{SS} \rangle$,
to investigate the possiblity of 4q states.
Shown in the middle panel of Fig.\ref{fig:borelfig}
are the effective masses
for $E_{th}$=1.0, 1.2, and 1.4 GeV.
We select the $E_{th}$=1.2 GeV case
and evaluate the mass as $\sim$0.90 GeV,
which is obviously lower than that of the 2q scalar meson case, 
$\sim$1.10 GeV 
in large $N_c$ limit,
and thus is considered as the mass of 4q state.
Although these absolute values depend on the
details of condensate values,
the inequality $m_{\rho}<m_{4q}<m_{S}$ is
insensitive to them.
In the right panel of Fig.\ref{fig:borelfig},
we show the $\langle \bar{q}q \rangle$
and $m_0^2$ dependence of
2q vector, scalar meson masses,
and of the 4q mass deduced from 
$\langle J_{VV}J^{\dag}_{SS} \rangle$. 

We have also investigated both $O(N_c^2)$ and $O(N_c)$
parts of $\langle J_{SS}J^{\dag}_{SS} \rangle$
($\langle J_{VV}J^{\dag}_{VV} \rangle$).
As for $O(N_c^2)$ part,
low energy contribution
is almost absent below two meson thresholds.
More precisely speaking,
acceptable effective mass plots can be obtained
only if we take much larger $E_{th}$ value than the two meson thresholds,
giving large effective masses.
This indicates that $O(N_c^2)$ part includes
only free scattering states
consistently with $1/N_c$ arguments.
As for $O(N_c)$ part,
$\langle J_{SS}J^{\dag}_{SS} \rangle$
($\langle J_{VV}J^{\dag}_{VV} \rangle$)
give similar values $\sim 0.9\ (0.8)$ GeV,
as $\langle J_{VV}J^{\dag}_{SS} \rangle$ case,
despite of the possible factorization violation
in the formers.

We have also investigated 4q correlators with $I=2$.
They do not allow us to find any stable effective mass plots or
any reasonably wide $M^2$ window.
This would indicate the absence of 4q pole in $I=2$ channel,
consistenly with experimental observations.

In conclusion, it is found that
the $1/N_c$ expansion
is very useful to analyze the leading $N_c$ properties of
multiquark states with separating scattering backgrounds.
The QSR results show consistencies
with expectations from $1/N_c$ arguments, 
and further provide the prediction 
on 4q components in the $\sigma$ meson
without $\pi\pi$ correlations.
The leading $N_c$ analyses 
for multiquark correlators may be also useful
to distinguish two meson molecules and tetraquarks,
which is discussed for the recently found
charmed mesons (X,Y,Z).
Since initial states are controled by interpolating fields,
and $1/N_c$ countings give observations on intermediate states,
the combination of them enables to separate the molecular type 
contributions in the spectral functions.
Further discussions will be made elsewhere.

T.K. thanks the organizers of {\it HSQCD}
for their hospitality during the workshop.
This work is supported by RIKEN, Brookhaven National
Laboratory and the U. S. Department of Energy 
[Contract No. DE-AC02-98CH10886], and by the 
Grant for Scientific Research (No. 20028004)
in Japan.

\end{document}